\documentclass[preprint,review,12pt]{elsarticle}

\usepackage{graphicx}
\usepackage{bm}
\usepackage{amsmath}
\usepackage{amssymb}
\usepackage{hyperref}
\usepackage{lineno}
\usepackage{natbib}

\usepackage{lineno}

\journal{Carbon}

\begin{document}

\begin{frontmatter}



\title{Alignment of graphene nanoribbons by an electric-field}


\author{Zhao Wang\corref{cor1}}
\cortext[cor1]{Corresponding author. Fax: +41 33 228 44 90. E-mail address: wzzhao@yahoo.fr (Z. Wang)}

\address{EMPA - Swiss Federal Laboratories for Materials Testing and Research, Feuerwerkerstrasse 39, CH-3602 Thun, Switzerland}

\begin{abstract}
In this paper, we develop an analytical approach to predict the field-induced alignment of cantilevered graphene nanoribbons. This approach is validated through molecular simulations using a constitutive atomic electrostatic model. Our results reveal that graphene's field-oriented bending angle is roughly proportional to the square of field strength or the graphene length for small deformations, while is roughly independent of graphene width. The effective bending stiffness and the longitudinal polarizability are also found to be approximately proportional to the square of graphene length. Compared with carbon nanotubes, graphene nanoribbons are found to be more mechanically sensitive to an external electric field.
\end{abstract}

\end{frontmatter}

\linenumbers
\section{Introduction}
Graphene's electronic gap tunable in external electromagnetic fields \cite{Novoselov2004,Zhang2005,Castro2007,Novikov2007a} makes it promising for a number of potential applications in nanoelectronic devices \cite{Bunch2007,Oostinga2008,Son2006a}. Since graphene is often supposed to work in a transverse electric field in such devices and its electronic transport properties strongly depend on its atomic structure \cite{Morozov2008,Duplock2004,Wakabayashi2009}, understanding of graphene's mechanical behaviors in an electric field is of great importance for nanoelectromechanical systems based on graphene. However, as a novel issue, the mechanical response of graphene to an external field has not yet been reported up to date. \textit{How does graphene deform in response to applied electric fields?} To answer this question, we developed a simple model to predict the field-induced alignment of cantilevered graphene nanoribbons (GNRs), demonstrating the coupling between the graphene's field-induced bending, molecular stiffness and electric polarization. This model is validated through molecular simulations, in view of the difficulties of experimental quantification of this electromechanical effects in nanoscale.

\begin{figure}[ht]
\centerline{\includegraphics[width=13cm]{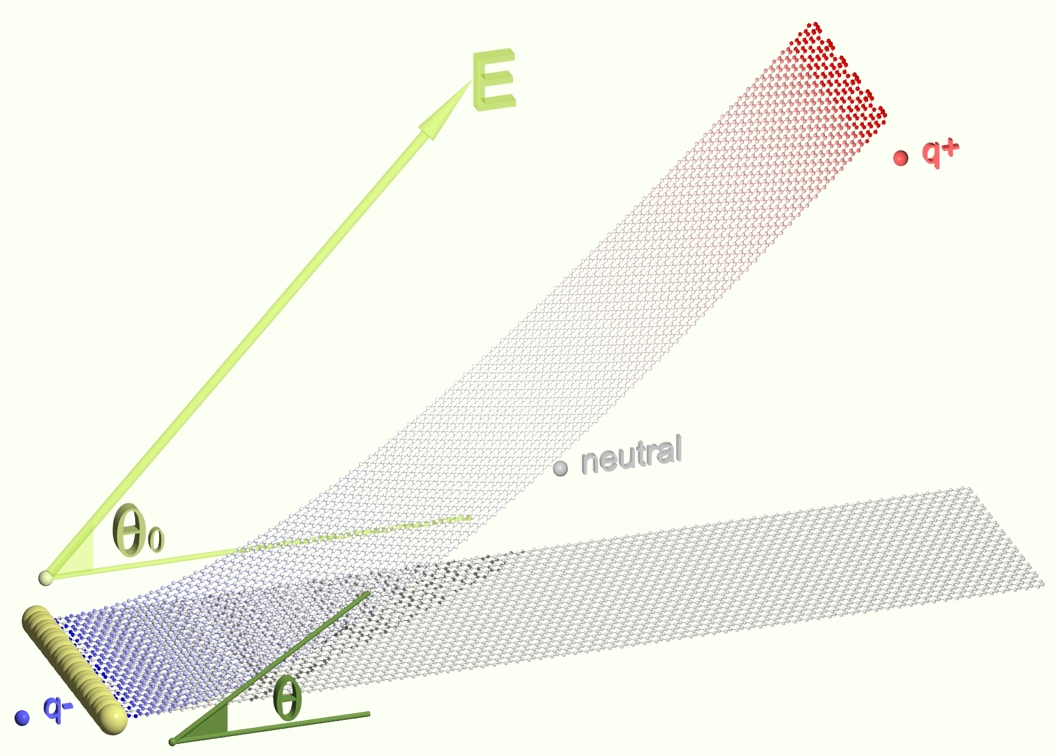}}
\caption{\label{fig:edeformations}
Alignment of a cantilevered GNR ($L\approx20$nm, periodic in width direction) to an electric field $E=0.3$V/nm. The color scale of atoms represents the intensity of induced charges $q$.
}
\end{figure}

If a thin nanostructure is brought into an electric field, electric polarization effects will induce a moment of force, which tends to orient the nanostructure toward the field direction \cite{joselevich-02}. This moment will make the nanostructure bent if one end of the nanostructure is fixed on a substrate (see Fig. \ref{fig:edeformations}). This alignment has been observed on carbon nanotubes (CNTs) in early experiments \cite{Poncharal-99}, and then exploited in designing nanorelays \cite{Kinaret2003a} and field emission devices \cite{purcell-02}. As can be expected in view of the large similarity in their atomic structures and mechanical properties \cite{Lee2008}, graphene nanoribbons (GNRs) should exhibit similar mechanical behaviors in an electric field. Furthermore, unlike CNTs, graphene exhibits strong mechanical anisotropy in its transverse direction, with a lateral stiffness about 30 times lower than that in the longitudinal direction \cite{Bosak2007,Lu1997}. This high lateral structural flexibility makes GNRs ideal field-sensing materials in resonators \cite{Bunch2007}, transistors \cite{Meric2008} or sensors \cite{Ang2008}. 

\section{Simulations}
In this work, molecular simulations were performed to compute equilibrium structures of cantilevered GNRs in an electric field, using an energy optimization method (full details of the simulation techniques can be found elsewhere \cite{zhaowang-07-03,Zhao1997}). The principle of this method is to minimize the potential energy of each atom, which consists of an internal potential due to the C-C chemical bonds and the long-range interaction, and of an external potential induced by applied electrostatic fields. The internal potential is calculated using the adaptive interatomic reactive empirical bond order (AIREBO) potential function \cite{Stuart2000a}, which has been used in recent studies on mechanical properties of CNTs \cite{Ni2002} and GNRs \cite{Shenoy2008}. The external potential describes the electrostatic interaction between the charges, the dipoles and the external field, it is computed using an atomic charge-dipole interacting model \cite{mayer-07-01,wang-09-02,langlet2006}, which has been validated through charge-injection experiments using an atomic force microscope \cite{wang-08-01}.

\section{Modeling}
The main reason of the field-induced alignment of a GNR is the effect of electric polarization, by which a quantity of positive and negative charges are shifted to opposite directions in graphene (see Fig. \ref{fig:edeformations}). A moment of a force pair arises from the electrostatic interactions between the field and the induced charges, and makes the graphene bent into the field direction. This field-driving moment highly correlated with the GNR's polarizability is resisted by the mechanical lateral stiffness of graphene, due to the  repulsive interactions between $\pi$ electrons and the rotation of $\sigma$ bonds. Considering the correlations between the electric polarizability, the bending stiffness, and the geometry of GNRs, the curvature $\omega$ of a cantilevered GNR in an electric field can be calculated as follows:

\begin{equation}
\label{eq:2}
\omega = \frac{M}{K} =\frac{ E^{2} (\alpha_{//} - \alpha_{\bot}) \sin \left[ 2(\theta_{0}-\theta) \right]} {2K}
\end{equation}

where $M$ is the bending moment induced by the electric polarization \cite{kozinsky-06}, $K$ denotes the effective bending stiffness of GNR, $L$ stands for the graphene length, $\theta$ is the deflection angle of GNR, $E=\left| \bm{E} \right|$, and $\theta_{0}$ represent the strength and the direction of the electric field, respectively, $\alpha_{//}$ and $\alpha_{\bot}$  stand for the longitudinal and the transverse molecular polarizabilities of GNR, respectively. $\theta$ is defined as the angle between the initial axis of graphene and the vector from one graphene end to another after deformation. Since the curvature $\omega$ can be approximated as $\omega = 2 \theta /L$ and usually $\alpha_{//}>>\alpha_{\bot}$, the governing equation can therefore be written as

\begin{equation}
\label{eq:3}
\theta \approx \frac{ L E^{2} \alpha_{//}^{*} \sin \left[ 2(\theta_{0}-\theta) \right] }  {4K^{*}}
\end{equation}

\begin{figure}[ht]
\centerline{\includegraphics[width=12cm]{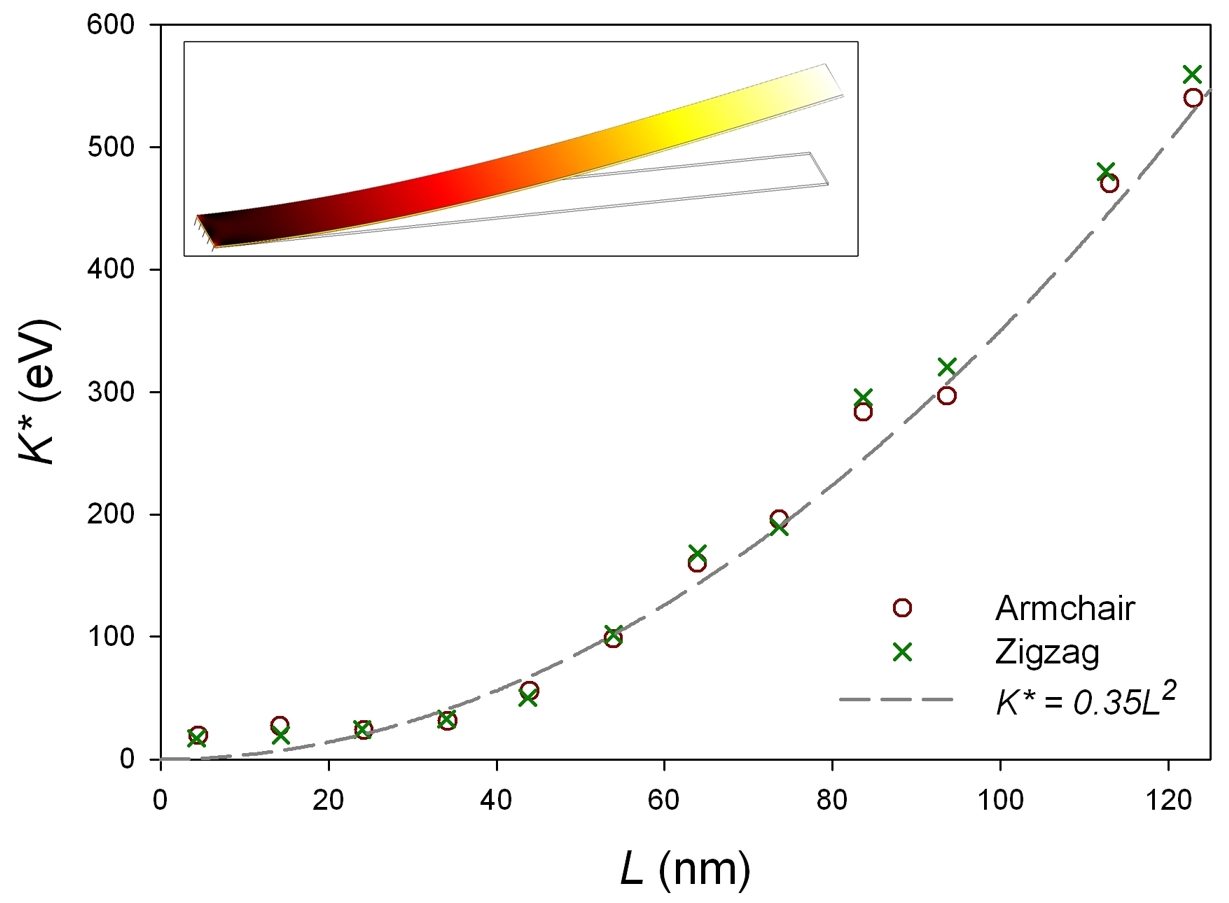}}
\caption{\label{fig:Bendingstiffness}
$K^{*}$ \textit{vs} $L$ from simulations, in which the periodic condition was applied in width direction. Symbols show simulation data and the dashed line stands for a fitting curve. The inset shows the main stress distribution in a curved graphene.}
\end{figure}

where $\alpha_{//}^{*}$ and $K^{*}$ stand for the quantities of $\alpha_{//}$ and $K$ per unit width, respectively. In Eq. \ref{eq:3}, the parameter representing the mechanical resistance of a GNR is the effective bending stiffness $K$. Here its values were directly computed from atomic simulations by applying a mechanical force at the free end of the GNRs in absence of electric field , instead of using a conventional formula for macroscopic continuous media as the product of the elastic modulus and the area moment of inertia, in order to avoid the problem of definition of the wall-thickness of a one-single-atom thick layer \cite{Huang2006a}. Note that previous studies showed that the bending stiffness of a CNT is an \textit{independent parameter} not necessarily related to the representative thickness by the classic formula \cite{Ru2000}. Our simulation results show that $K^{*}$ is about $20$eV when $L<20$nm and is roughly proportional to $L^{2}$ when GNRs get longer (see Fig. \ref{fig:Bendingstiffness}). These simulation data can be fitted using a simple equation as

\begin{equation}
\label{eq:k}
K^{*}=A L^{2} \,\,\,\,\,\,\,\,\,\, (L>20nm)
\end{equation}

where $A=0.35$eV$\cdot$nm$^{-2}$ for GNRs with either armchair or zigzag edges. For comparison, we also calculated the effective bending stiffness of CNTs. It is found that the value of $K$ of a GNR ($L \approx 10$nm, $K \approx 19.2 $eV$\cdot$nm) is about 20 times smaller than that of a (5, 5) CNTs of the same length ($K \approx 570$eV$\cdot$nm). This large difference in the transverse (lateral) stiffness implies that the alignment of GNRs can be much more significant than that of CNTs for a given magnitude of electric polarization. 

\begin{figure}[htp]
\centerline{\includegraphics[width=12cm]{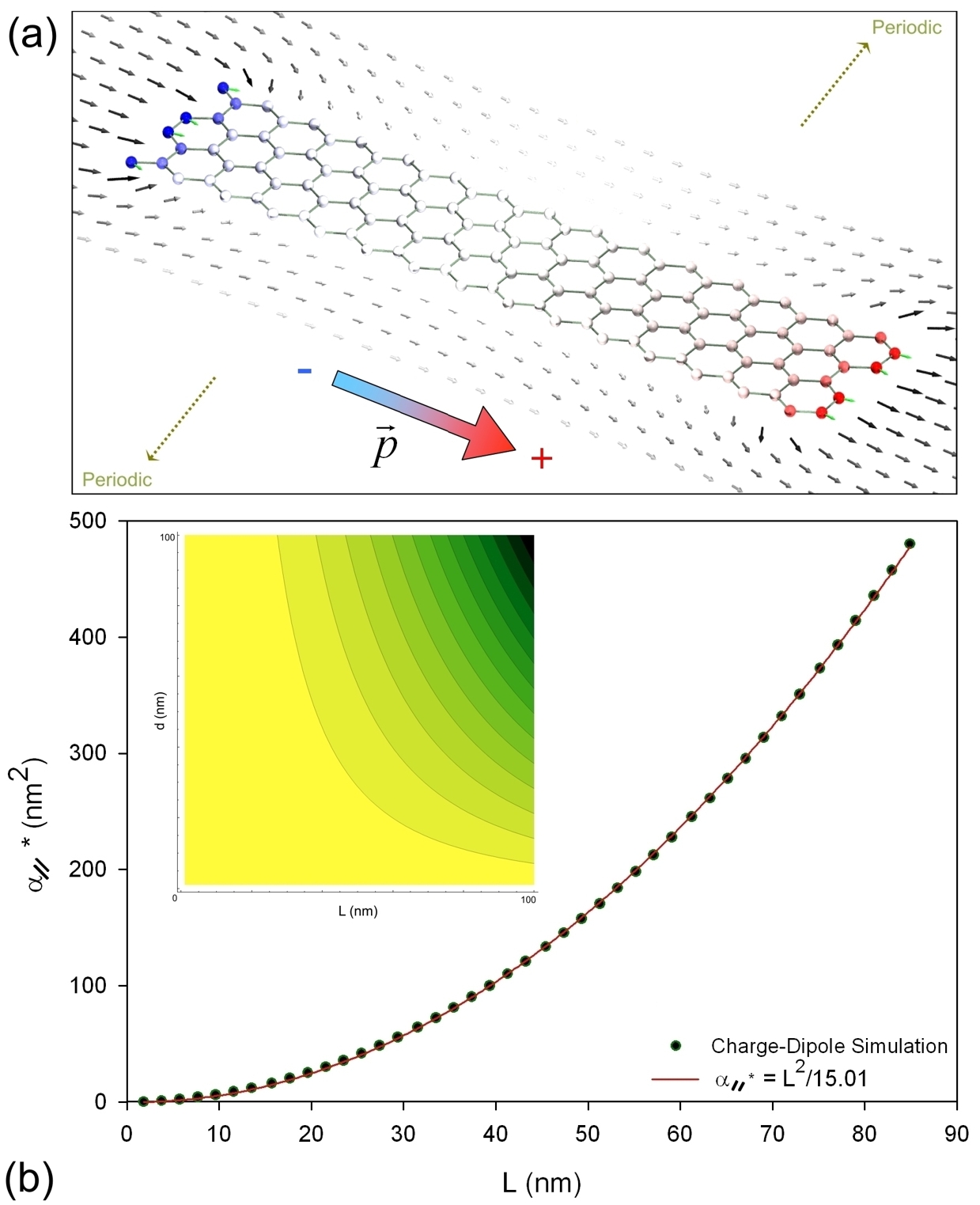}}
\caption{\label{fig:polarizability}
(a) Polarization of a GNR in a longitudinal electric field. Color scale of atom represents the intensity of induced charges. Dark arrows show the electric field around the graphene. (b) Longitudinal polarizability per $1$nm width $\alpha_{//}^{*}$ \textit{vs} $L$. Circles denote simulation data and the dashed line stands for a numerically fitting curve. Coupled dependence of $\alpha_{//}$ on $L$ and width $d$ is shown in the inset, in which the color scale represents the amplitude of $\alpha_{//}$. 
}
\end{figure}

Another important parameter in Eq. \ref{eq:3} is $\alpha_{//}^{*}$. Its value was determined using electrostatic simulations based on the atomic charge-dipole model \cite{mayer-07-01}. In these simulations, the value of $\alpha_{//}^{*}$ was calculated from the definition $\vec{p}=\vec{E} \bar{\bar{\alpha}}$, where $\bar{\bar{\alpha}}$ is the the matrix of molecular polarizability, $\vec{p}$ and $\vec{E}$ stand for the vectors of the induced molecular dipole (see Fig. \ref{fig:polarizability} (a)) and the applied electric field, respectively. As shown in Fig. \ref{fig:polarizability} (b), our simulation results suggest that $\alpha_{//}^{*}$ is roughly proportional to $L^{2}$.

\begin{equation}
\label{eq:a}
\alpha_{//}^{*} =\frac{L^{2}}{B} \,\,\,\,\,\,\,\,\,\, (L>6nm)
\end{equation}

where $B = 15.01$ is a constant for either armchair or zigzag GNRs, since no large difference has been found between $\alpha_{//}^{*}$ of these two types of graphene. Since Eq. \ref{eq:a} shows a typical metallic behavior of graphene, we note that $\alpha_{//}^{*}$ of semi-conducting graphene (minimum lateral dimension $< 6$nm \cite{Ritter2009}) should hold a linear relationship with $L$. Putting the empirical fits to $\alpha_{//}$ and $K^{*}$ into Eq. \ref{eq:3}, we finally obtain the governing equation of the electrostatic alignment of GNRs as follows:

\begin{equation}
\label{eq:6}
\theta = \frac{E^{2} L \sin \left[ 2(\theta_{0}-\theta) \right] }  {C}   \,\,\,\,\,\,\,\,\,\, (L>20nm)
\end{equation}

where $C=4AB\approx21$eV$\cdot$nm$^{-2}$. We note that, since the geometry periodic condition was applied in the width direction in our calculations, we effectively simulated graphene of infinite width. Thus, the edge disorder effects on structural \cite{Fasolino2007} and mechanical properties \cite{Shenoy2008} of graphene were neglected. 

\begin{figure}[htp]
\centerline{\includegraphics[width=12cm]{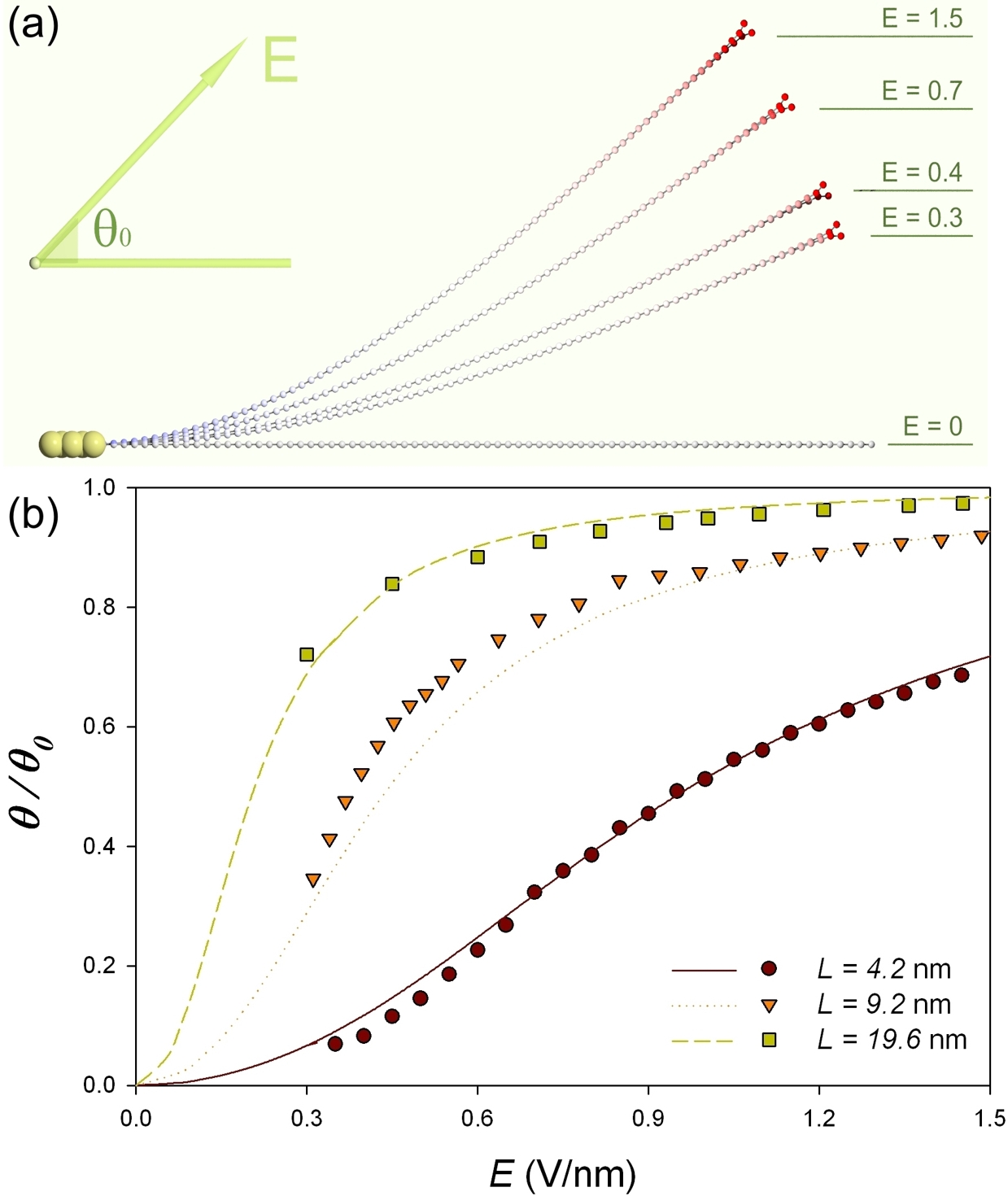}}
\caption{\label{fig:FieldStrength}
(a) Cross sections of a GNR ($L\approx9.2$nm) in different electric fields $\bm{E}$ ($\theta_{0}=\pi/4$), (a video recording file is available on line). (b) Alignment ratio $\theta/\theta_{0}$ \textit{versus} the field strength $E$ for GNRs of different lengths. The symbols represent numerical data and the lines stand for results predicted by analytical model. 
}
\end{figure}

\section{Results and discussions}
We can see from Eq. \ref{eq:6} that, in small deformation region, $\theta$ is roughly proportional to $E^{2}$, $\sin{\theta_{0}}$ or $L$. In Fig. \ref{fig:FieldStrength}, we plot data of the alignment ratio $\theta/\theta_{0}$ as a function of $E$ from molecular simulations in which periodic geometry condition is applied in width direction. A quantitative agreement was obtained between the simulation data and the analytical prediction using Eq. \ref{eq:6}. The main inaccuracy can be considered to be contributed from the geometry approximations and the possible slight change of polarizability due to the curvature of graphene. We can see that the S-shaped curve of $\theta/\theta_{0}$ tends to be flat near the maximal value $1$ when the GNRs are well aligned to the field direction. In such a case that $\sin \left[ 2(\theta_{0}-\theta) \right] \approx 2(\theta_{0}-\theta)$, Eq. \ref{eq:6} can be simplified to 

\begin{equation}
\label{eq:7}
\frac{\theta }{\theta_{0}} \approx 1 - \frac{C}{C+ 2 E^{2} L}
\end{equation}

for large deformation. Compared with GNRs in same sizes, CNTs were found to be less flexible in an electric field, e.g., a (5,5) SWCNT ($L \approx 10$nm) can be bent to $\theta/\theta_{0} \approx 0.4$ in an electric field ($E \approx 2.0$V/nm and $\theta_{0}=\pi/4$), while required field strength for producing the same amplitude of alignment for a GNR with the same width and length is about 7 times smaller ($E \approx 0.3$V/nm). Analysis on the amplitudes of induced polarization in a GNR and a CNT shows that this difference is mainly due to the fact that the bending stiffness of the GNR is much lower than that of the CNT. Furthermore, we can predict that a multi-layered graphene should be less sensitive to an electric field than a single-layered one is, because of the electric screening effects and the friction between the layers \cite{zhaowang-07-03}.  

\section{Conclusion}
In conclusion, we have developed an analytical model to predict the alignment of cantilevered GNRs to an electrostatic field. Parameters used in this model such as the polarizability and the bending stiffness were determined from numerical fits to the data of simulations using an atomic electrostatic charge-dipole model and an empirical pseudo-chemical potential. This model showed that the alignment angle roughly follows a linear relationship with the square of field strength and the graphene length when the deformation remains small. It was also found that, for GNRs with either armchair or zigzag edges, their effective bending stiffness and longitudinal polarizability are both approximately proportional to the square of graphene length. Comparison showed that a GNR can be more easily oriented to electric fields than a CNT does, due to the GNR's low transverse bending stiffness. 

\section*{Acknowledgments}
We gratefully thank S. J. Stuart and R. Langlet for help with the numerics. D. Stewart, A. Mayer, M. Devel and W. Ren are acknowledged for fruitful discussions.

\end{document}